# Chemical Bonding in the Boron Buckyball

Arnout Ceulemans,* Jules Tshishimbi Muya, G. Gopakumar, and Minh Tho Nguyen

*Department of Chemistry, and Institute for Nanoscale Physics and Chemistry (INPAC), University of Leuven, Celestijnenelaan 200F, B-3001 Leuven, Belgium*

According to quantumchemical calculations the boron buckyball, $B_{80}$, is very similar to the buckminsterfullerene carbon analogue. Both allotropes have a truncated icosahedral structure, which in the case of boron is complemented by an additional set of 20 boron atoms, capping the 20 hexagonal faces.[1] The HOMO and LUMO of icosahedral $B_{80}$ have the same symmetry as their counterparts in $C_{60}$, and moreover very similar shapes (Fig. 1). On the other hand the perfectly symmetrical $I_h$ structure of $B_{80}$ is not a local minimum, but relaxes to a slightly puckered cage with $T_h$ symmetry.[2] In a recent contribution Prasad and Jemmis argue that $B_{80}$ and $C_{60}$ are *isoelectronic*, both having 240 valence electrons.[3] This implies that the 60 electrons on the caps are transferred to the 60 vertices of the truncated icosahedron, to make up for the electron deficiency of boron versus carbon. Hence:

$$B_{80} \rightarrow B_{60}^{-60} \rightarrow C_{60} \quad (1)$$

In order to claim the same bonding pattern for both buckyballs, not only the total electron counts must be equal, but the *symmetries* of the occupied valence orbitals also must show a perfect match. Here we analyze the chemical bonding in $B_{80}$ and indeed demonstrate the symmetry correspondence of the occupied orbitals. Furthermore we identify the electron transfer channels between the caps and the frame.

An orbit of a group is defined as a set of elements, which are permuted among themselves by the symmetry operations of the group. In the case of $B_{80}$ the 80 atoms form two separate orbits, which we will denote as $O_{60}$ and $O_{20}$. The sixty-atom orbit consists of the boron atoms at the vertices of the truncated icosahedral buckyball, and the twenty-atom orbit comprises the caps, which are situated at the vertices of a dodecahedron.

The valence shell of boron consists of 2s and 2p orbitals. The 2p orbitals on a spherical cluster like $B_{80}$ form a radial component and two tangential components, which we will denote as $p_\sigma$ and $p_\pi$ respectively. The irreducible symmetry representations of the molecular orbitals spanned by these components may easily be derived using induction theory.[4] In the cluster 2s and $2p_\sigma$ are both σ-type orbitals and thus span the same symmetries. One has:

$$\Gamma_\sigma^{60} = A_g + T_{1g} + 2T_{1u} + T_{2g} + 2T_{2u} + 2G_g + 2G_u + 3H_g + 2H_u$$
$$\Gamma_\pi^{60} = A_u + 3T_{1g} + 3T_{1u} + 3T_{2g} + 3T_{2u} + 4G_g + 4G_u + 5H_g + 5H_u \quad (2)$$
$$\Gamma_\sigma^{20} = A_g + T_{1u} + T_{2u} + G_g + G_u + H_g$$
$$\Gamma_\pi^{20} = T_{1g} + T_{1u} + T_{2g} + T_{2u} + G_g + G_u + 2H_g + 2H_u$$

A further useful relation, which may be obtained from this equation reads:

$$\Gamma_\sigma^{20} + \Gamma_\pi^{20} = \Gamma_\sigma^{60} \quad (3)$$

Symmetry elements also will permute localized chemical bonds, which correspond to pair wise interactions between atoms. Bonds too will thus form orbits. The 90 edge bonds in buckminster fullerene constitute two orbits: $O_{60} + O_{30}$. The $O_{60}$ orbit contains the 5-6 bonds adjacent to a pentagon and a hexagon, while the $O_{30}$ orbit is the set of the 6-6 bonds adjacent to two hexagons. The σ-bonds along the 5-6 edges span exactly the same $\Gamma_\sigma^{60}$ symmetry as the 2s and $2p_\sigma$ orbitals on the $O_{60}$ orbit. The 30 σ-bonds along the 6-6 edges transform as $\Gamma_\sigma^{30}$:

$$\Gamma_\sigma^{30} = A_g + T_{1u} + T_{2u} + G_g + G_u + 2H_g + H_u \quad (4)$$

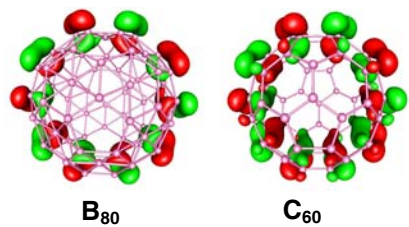

**Figure 1**. Shape of the highest occupied molecular orbitals of $B_{80}$ and $C_{60}$.

The site symmetry in the centre of a 6-6 edge is $C_{2v}$. Eq. 4 lists all irreducible representations of objects which are totally symmetric in this site group, i.e. which transform as the $a_1$ representation in $C_{2v}$. For the transformation properties of anti-bonds localized along these edges, one needs to consider objects that are of $b_1$-type in $C_{2v}$. The induced representation of the latter set is:

$$\Gamma_{\sigma^*}^{30} = T_{1g} + T_{1u} + T_{2g} + T_{2u} + G_g + G_u + H_g + H_u \quad (5)$$

The cap boron atoms are at the vertices of a dodecahedron, the edges of which are perpendicular to the 6-6 bonds. The number of vertices in such a trivalent cage is equal to twice the number of edges. This number relation has a symmetry extension[5] in terms of irreducible representations:

$$\Gamma_\sigma^{20} + \Gamma_\pi^{20} = \Gamma_\sigma^{30} + \Gamma_{\sigma^*}^{30} \quad (6)$$

The chemical interpretation of eq. 6 is straightforward. If one forms with the cap orbitals three hybrids directed towards the centres of the 6-6 bonds, one can construct a two-centre bond and anti-bond along each edge. Note that for the $C_{2v}$ sites perpendicular and parallel anti-bonds induce the same $I_h$ representations.

In spite of its spherical nature the chemical bonding in $C_{60}$ may simply be described in terms of a localized valence bond theory. The $2s+2p_\pi$ orbitals on carbon are $sp^2$ hybridized and provide a fully bonding network along the 90 edges of the buckyball. This accounts for 180 electrons. In addition on each carbon there is one radial $p_\sigma$ orbital containing one electron. These orbitals form a conjugated π-bonding network which is mainly localized along the 6-6 edges, and contains 60 electrons.[6] In total the 240 electrons of $C_{60}$ occupy 120 orbitals with the following symmetries:

$$\Gamma_{bond}(C_{60}) = \Gamma_\sigma^{60} + 2\Gamma_\sigma^{30} \quad (7)$$

From eqs. (2) and (4) the symmetries of the occupied orbitals of the valence shell of $C_{60}$ are thus identified as follows:

$$\Gamma_{bond}(C_{60}) = 3A_g + T_{1g} + 4T_{1u} + T_{2g} + 4T_{2u} \quad (8)$$
$$+ 4G_g + 4G_u + 7H_g + 4H_u$$

In this model scheme π-conjugation in $C_{60}$ is of the alternant type. In reality double bonding is delocalized over all 90 bonds, although 6-6 bonds show more double bonding character than 5-6 bonds, as is evidenced by bond lengths 1.40 versus 1.46 Å respectively.

In table 1 we list the symmetries and energies of valence orbitals of $B_{80}$ at B3LYP/SV(P) level calculated using TURBOMOLE quantum chemistry program.[7]

**Table 1**. Valence orbitals of $B_{80}$. ($6h_u$ is the HOMO)

| mos | Energy (e.V) | mos | Energy (e.V) | mos | Energy (e.V) |
|---|---|---|---|---|---|
| $3a_g$ | -21.968 | $2t_{1g}$ | -13.837 | $6g_g$ | -8.664 |
| $4t_{1u}$ | -21.593 | $5g_u$ | -13.06 | $5h_u$ | -8.085 |
| $5h_g$ | -20.889 | $4h_u$ | -12.649 | $7t_{2u}$ | -8.049 |
| $4g_u$ | -20.161 | $4a_g$ | -12.157 | $7g_u$ | -7.607 |
| $4t_{2u}$ | -19.096 | $6t_{2u}$ | -12.115 | $10h_g$ | -7.5 |
| $4g_g$ | -18.776 | $2t_{2g}$ | -12.005 | $7g_g$ | -7.312 |
| $6h_g$ | -17.728 | $5a_g$ | -11.726 | $11h_g$ | -6.497 |
| $3h_u$ | -17.166 | $6t_{1u}$ | -11.263 | $6h_u$ | -5.633 |
| $5t_{2u}$ | -16.091 | $8h_g$ | -11.147 | $8t_{1u}$ | -3.711 |
| $5t_{1u}$ | -14.972 | $7t_{1u}$ | -11.102 | $3t_{1g}$ | -3.197 |
| $5g_g$ | -14.940 | $9h_g$ | -10.217 | $8t_{2u}$ | -3.031 |
| $7h_g$ | -14.376 | $6g_u$ | -9.226 | | |

One verifies that the symmetries exactly match the valence shell of $C_{60}$, as listed in eq. (8), which confirms that both clusters are isoelectronic. Upon more detailed examination the valence shell consists of two sub-bands, representing the 90 σ-bonds and 30 π-bonds. These sub-bands overlap quite extensively: in the region between –14 to –8 eV they are strongly hybridized and difficult to assign. Nonetheless based on the dominant component the sequence inside the sub-bands can be analyzed on the basis of spherical parentages.

For the 90 σ-orbitals one can retrieve in ascending order of energy the complete spherical shells from L=0 till L=8, which account for 81 orbitals. The remaining 9 orbitals transform as $G_u+H_u$ belonging to L=9. In table 2 we identify these levels in terms of the MO's. The 30 π-orbitals form a separate spherical sequence, with complete shells from L=0 to L=4, accounting for 25 of the 30 orbitals, the remaining 5 corresponding to the $H_u$ HOMO of the buckyball, belonging to the L=5 shell.

From this detailed analysis it is thus clear that the bonding orbitals in $B_{80}$ precisely mimic the structure of the valence shell in $C_{60}$. The optimized bond lengths also reflect the alternant scheme, with 6-6 and 5-6 bond lengths of 1.67 and 1.74 Å respectively. On the other hand $B_{80}$ cannot really be described as a $B_{60}$ cage with 60 negative charges. In fact the NBO population analysis, carried out using Gaussian 03 program,[8] showed only a very small charge transfer of ~3 electrons from $O_{20}$ to $O_{60}$. We thus must determine how the 60 electrons of the caps flow into the buckyball.

For the cap atoms to participate in the bonding it is not necessary that their electrons be fully transferred to the $B_{60}$ frame. What is certainly required is a symmetry allowed overlap between the orbitals of the cap atoms and the bonding orbitals of the frame. The actual degree of charge transfer then depends on the relative weights of the $O_{20}$ versus $O_{60}$ orbits in the resulting combination. But in order to obtain a consistent bonding picture for the entire cluster, a stronger requirement also is needed: the overlap argument should not only apply to individual levels, but to all levels of a given orbit.

**Table 2**. Spherical shells and σ- and π-band of $B_{80}$

| L | $I_h$ | σ-band | π-band |
|---|---|---|---|
| 0 | $A_g$ | $3a_g$ | $5a_g$ |
| 1 | $T_{1u}$ | $4t_{1u}$ | $7t_{1u}$ |
| 2 | $H_g$ | $5h_g$ | $10h_g$ |
| 3 | $T_{2u} + G_u$ | $4t_{2u} + 4g_u$ | $7t_{2u}+7g_u$ |
| 4 | $G_g + H_g$ | $4g_g + 6h_g$ | $7g_g+11h_g$ |
| 5 | $T_{1u} + T_{2u} + H_u$ | $5t_{1u} + 5t_{2u} + 3h_u$ | $6h_u$ |
| 6 | $A_g + T_{1g} + G_g + H_g$ | $4a_g + 2t_{1g} + 5g_g + 7h_g$ | |
| 7 | $T_{1u} + T_{2u} + G_u + H_u$ | $6t_{1u} + 6t_{2u} + 5g_u + 4h_u$ | |
| 8 | $T_{2g} + G_g + 2H_g$ | $2t_{2g} + 6g_g + 8h_g + 9h_g$ | |
| 9 | $T_{1u} + T_{2u} + 2G_u + H_u$ | $6g_u + 5h_u$ | |

In order to understand the bonding role of the caps, one thus must consider the $\Gamma_\sigma^{20}+\Gamma_\pi^{20}$ symmetries, which describe the σ+π valence shell of the cap atoms, and verify how they are partitioned in orbits. The two possible partitions are described in eqs. (3) and (6): cap boron atoms can either participate in all sixty 5-6 bonds via the $\Gamma_\sigma^{60}$ channel, or can provide 30 orbital levels transforming as $\Gamma_\sigma^{30}$. Analysis of the orbital composition in the quantum chemical results, as given in the supplementary information, shows that the cap atoms contribute nearly 30 orbitals, the symmetries of which are exactly the ones contained in $\Gamma_\sigma^{30}$. Hence the second option is the correct one. The remaining cap orbitals transforming as $\Gamma_{\sigma*}^{30}$ will remain virtual. Upon closer inspection it is noticed that the bonding contribution of the cap atoms is mostly concentrated in the σ-band, while the π-band close to the Fermi level is more localized on the frame orbit.

While the present local bonding picture based on whole orbits of the icosahedral group is very useful to explain the main features of the bonding, it cannot account for the $T_h$ puckering that is revealed by the calculations. Clearly local bonding motifs forming $T_h$ orbits must then be involved. One possible candidate is a hexagon, surrounded by the star of its three neighboring hexagons.

Finally for *m* the number of hexagons in a $C_n$ fullerene, the hexagon-capped boron cage, $B_{n+m}$, will be isoelectronic for *m=n*/3. Since for a fullerene one also has *m*=0.5*n*-10, the unique fully hexagon capped isoelectronic cage will be $B_{80}$.

**Acknowledgment** The authors are indebted to the Flemish Science Fund (FWO-Vlaanderen) and the KU Leuven Research Council for continuing financial support.

**Supporting Information Available:** Contribution of 2s and 2p orbitals of cap atoms for different molecular orbitals. This material is available free of charge via the Internet at http://pubs.acs.org.

**REFERENCES**

(1) N. G. Szwacki, A. Sadrzadeh, B. I. Yakobson, *Phys. Rev. Lett.* 2007, 98; 166804.
(2) G. Gopakumar, M. T. Nguyen, and A. Ceulemans, *Chem.Phys.Lett* 2008; 450,175
(3) D. L. V. K. Prasad and E. D. Jemmis, *Phys. Rev. Lett.* 2008, 100, 165504
(4) P. W. Fowler and C. M. Quinn, *Theor Chim Acta*. 1986; 70, 333-350
(5) A. Ceulemans and P.W.Fowler, *Nature*. 1991; 353, 6339
(6) P. W. Fowler and A. Ceulemans, *J. Phys. Chem.* 1995;99, 508-510
(7) R. Ahlrichs *et al. Chem. Phys. Lett.* 1989; 162, 165
(8) Gaussian 03, Rev. D.02, M. J. Frisch, *et al.*, Gaussian, Inc., Wallingford CT, 2004.

Graphic Entry

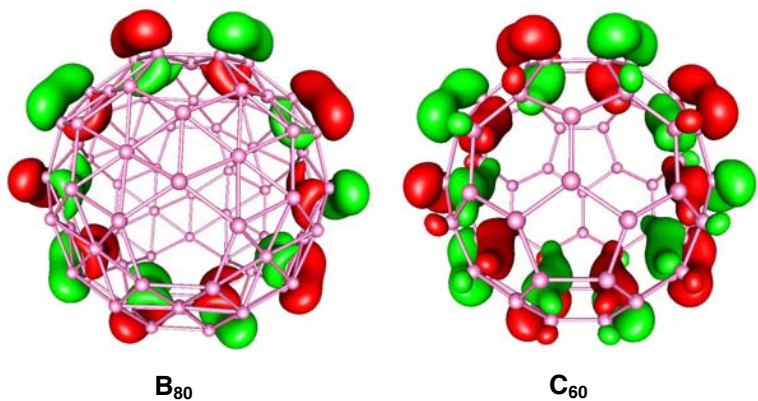

**B$_{80}$**  **C$_{60}$**

ABSTRACT: A symmetry analysis of the valence shell in the boron buckyball B$_{80,}$ shows a perfect match with the symmetry of the valence shell in C$_{60}$ .The cap atoms participate in the bonding by 30 orbitals which transform as the σ-bonds along the 6-6 edges. Far reaching consequences for chemical and physical properties of B$_{80}$, which until now only exists in *silico*, are expected.